\documentclass[aps,prl,twocolumn,amssymb,showpacs]{revtex4}

\usepackage{amsmath}

\def\shorteqn{\abovedisplayskip=-\baselineskip\equation}
\def\endshorteqn{\endequation}
\newenvironment{shorteqn*}{\shorteqn}{\nonumber\endshorteqn}
\renewcommand{\(}{\left(}
\renewcommand{\)}{\right)}

\renewcommand{\hbar}{\hslash}

\renewcommand{\>}{\rangle}
\newcommand{\half}{\tfrac{1}{2}}

\newcommand{\tr}{\operatorname{tr}}

\begin{document}
\title{One-qubit reduced states of a pure many-qubit state: polygon inequalities}

\author{ A.~Higuchi,$^1$ A.~Sudbery$^2$ and J.~Szulc$^3$}

\affiliation{Department of Mathematics, University of York, Heslington, York
YO10 5DD, UK\\
email: $^1$ah28@york.ac.uk, $^2$as2@york.ac.uk, $^3$js115@york.ac.uk}

\date{15 August 2002}

\begin{abstract}
We show that a necessary and sufficient condition for a set of $n$
one-qubit mixed states to be the reduced states of a pure $n$-qubit
state is that their smaller eigenvalues should satisfy polygon
inequalities: each of them must be no greater than the sum of the others.

\end{abstract}


\maketitle


In this paper we study the quantum analogue of the
marginal distributions of a joint probability distribution.
The results reveal some surprising aspects of
pure states of multipartite quantum systems. 

In a pure state of many particles, each subset $P$ of the particles is
in a mixed state $\rho_P$, the \emph{reduced state} of the subset $P$.
These reduced states are subject to conditions such as the
consistency conditions
\begin{equation}\tr_Q\rho_{P\cup Q} = \tr_R\rho_{P\cup R}
\label{consis}\end{equation}
for any subsets $P,Q,R$ such that $P$ and $Q$ are disjoint and $P$ and
$R$ are disjoint. However, equations of this
type do not exhaust the conditions: Linden, Popescu and Wootters
\cite{NoahSanduBill:marginals} have given an example of a set of three
two-qubit density matrices which satisfy all such conditions but 
cannot be the reduced two-qubit states
of a pure state of three qubits. Another example is given by the system
of four qubits, in which it is impossible for all three reduced
two-qubit states to be totally mixed \cite{bobcarol} though all
the consistency conditions \eqref{consis} are satisfied. The full set of
conditions is not known.

In this paper we completely determine the possible one-qubit reduced
states of a pure state of $n$ qubits. (There is no problem for mixed
states, since any $n$ one-qubit mixed states $\rho_1,\ldots\rho_n$
are the reduced states of the $n$-qubit mixed state
$\rho_1\otimes\cdots\otimes\rho_n$.) This is equivalent to determining
the set of possible eigenvalues for each qubit, i.e. the marginal
probability distributions. Brun and Cohen
\cite{BrunCohen:3qubits} have pointed out that in the case $n=3$ the
determinants of the reduced states must satisfy triangle inequalities,
but not every set of three $2\times 2$ density matrices satisfying these
inequalities is a possible set of reduced states of a pure three-qubit
state. We will show that such density matrices must satisfy the stronger condition that
their smaller eigenvalues satisfy triangle inequalities. For $n$ qubits,
a necessary condition is the obvious generalisation of the triangle inequalities:
\begin{equation}\label{polygonk}
\lambda_k \le \lambda_1 + \cdots + \lambda_{k-1} + \lambda_{k+1} +
\cdots + \lambda_n
\end{equation}
where the eigenvalues of the density matrix $\rho_k$ of the $k$th qubit
are $\lambda_k$ and $1 - \lambda_k$ with $\lambda_k \le 1/2$. These
inequalities completely
characterise the possible sets of eigenvalues, and therefore the
possible reduced one-qubit states, of a pure $n$-qubit state.

These results reveal some surprising aspects of pure quantum states. One
concerns the comparison with the classical notion of a pure state
in the sense of a probability distribution. The notion of a 
\emph{pure state} applies to both quantum and classical
systems, if we consider a state of a classical system to be a
probability distribution for the values of variables which actually have
precise values. Then a pure state of a classical system (defined to be an
extreme point of the convex set of probability distributions)
corresponds to perfect knowledge of the variables, so that all
probabilities are 0 or 1. A pure state of a quantum system corresponds
to maximum knowledge, but the characteristic feature of quantum mechanics
is that even in a pure state there are variables for which perfect knowledge
is not available; indeed, any possible probability distribution will
occur for some variable. However, given any pure state there is 
always some physical variable
for which the probabilities are all 0 or 1 (for example, the projection
operator onto the state). This is the quantum version
of the above characterisation of classical pure states. 

This analogy between quantum and classical pure
states disappears when we consider marginal states. Classically, a joint
distribution of many variables is pure if and only if each one-variable
marginal distribution is pure. We know that this is not true in quantum
mechanics; the fact that a pure multipartite state may give mixed
one-party states is the phenomenon of entanglement. Nevertheless, our results
place limits on the one-party states; not every probability distribution
is possible. 

The most surprising aspect of this is that it is a statement about an overall
property of the whole system, namely the purity of its state, which
depends only on \emph{local} measurements. We emphasize that this is
only a negative statement: from the knowledge that the one-party reduced
states satisfy the polygon inequalities one cannot normally deduce that
the multipartite state must be pure (this is only possible if the
one-party states are themselves pure); but the local information in a
violation of the polynomial inequalities does give the overall
information that the multipartite state cannot be pure.


To prove the assertion \eqref{polygonk}, we expand the pure $n$-qubit state $|\Psi\>$
in the Schmidt basis \cite{BrunCohen:3qubits, CohenBrun:Schmidt,
localinv}. Let $|\phi_0^{(k)}\>$ and $|\phi_1^{(k)}\>$ be the eigenstates of the reduced
state $\rho_k$ with eigenvalues $\lambda_k$ and $1 - \lambda_k$
respectively; we will show that 
\begin{equation}\label{polygon}
\lambda_1 \le \lambda_2 + \cdots + \lambda_n.
\end{equation}
We assume $\lambda_1 > 0$ (if not, the inequality \eqref{polygon} is
trivially satisfied).
We note that $|\phi_0^{(1)}\>$ and $|\phi_1^{(1)}\>$ are the states
of the first qubit occurring in the
Schmidt decomposition of $|\Psi\>$ when the system is divided into two
parts, one containing the first qubit and the other containing the rest, so 
\begin{equation}\label{Schmidt}
|\Psi\> = A|\phi_0^{(1)}\>|\Phi_0^{(1)}\> + B|\phi_1^{(1)}\>|\Phi_1^{(1)}\>
\end{equation}
where $|\Phi_0^{(1)}\>$ and $|\Phi_1^{(1)}\>$ are orthogonal
normalised $(n-1)$-qubit states, $A$ and $B$ are real and $A^2=
\lambda_1 = 1 - B^2$. Since $\lambda_1$ is the smaller eigenvalue,
$0<A\le B$. We write $|i_1\ldots i_n\> = |\phi_{i_1}^{(1)}\>\ldots|\phi_{i_n}^{(n)}\>$ and
expand $|\Psi\>$ in this basis:
\begin{align*}
|\Psi\rangle &= A \sum_{i_2,i_3,\ldots, i_{n}}
a_{i_2 i_3 \cdots i_{n}}|0 i_2 i_3 \cdots i_{n}\rangle\\  
&\quad + B \sum_{i_2, i_3,\ldots, i_{n}}
b_{i_2 i_3 \cdots i_{n}}|1 i_2 i_3 \cdots i_{n}\rangle
\end{align*}
with
\begin{equation}
\sum_{i_2, i_3,\ldots, i_{n}}
|a_{i_2 i_3 \cdots i_{n}}|^2
= \sum_{i_2, i_3,\ldots, i_{n}}
|b_{i_2 i_3 \cdots i_{n}}|^2 = 1\,.
\end{equation}
Then
\begin{align*}
\lambda_k &= A^2\sum_{i_2,i_3,\ldots,i_n; i_k=0}|a_{i_2i_3\cdots
i_n}|^2 \\
&\quad + B^2\sum_{i_2,i_3,\ldots,i_n; i_k=0}|b_{i_2i_3\cdots i_n}|^2
\end{align*}
so that 
\begin{align*}
\lambda_2 + \cdots + \lambda_n & =   
A^2 \sum_{i_2,i_3,\ldots, i_{n}} N_{i_2 i_3\cdots i_n}
|a_{i_2 i_3\cdots i_n}|^2\\
&\quad + B^2 \sum_{i_2,i_3,\ldots, i_{n}} N_{i_2 i_3\cdots i_n}
|b_{i_2 i_3\cdots i_n}|^2\,, 
\end{align*}
where 
\begin{equation}
N_{i_2 i_3 \cdots i_n} \equiv n-1-(i_2+i_3+\cdots + i_n)
\end{equation}
is the number of times ``0" appears in $(i_2,i_3,\ldots,i_n)$.  Hence
$N_{i_2 i_3 \cdots i_n} \geq 1$ if $(i_2,i_3,\ldots,i_n) \neq (1,1,\ldots,1)$.
This implies that
\begin{eqnarray}
\lambda_2 + \cdots + \lambda_n
& \geq  &  
A^2 \left(\sum_{i_2,i_3,\ldots, i_{n}}
|a_{i_2 i_3\cdots i_n}|^2 - |a_{11\cdots 1}|^2\right)\nonumber \\
&& + B^2 \left( \sum_{i_2,i_3,\ldots, i_{n}} 
|b_{i_2 i_3\cdots i_n}|^2 - |b_{11\cdots 1}|^2\right)\nonumber \\
& = & A^2 (1-|a_{11\cdots 1}|^2) + B^2(1-|b_{11\cdots 1}|^2)\nonumber \\
& \geq & A^2(2-|a_{11\cdots 1}|^2 - |b_{11\cdots 1}|^2)\,,\label{sum}
\end{eqnarray}
since $B^2 \geq A^2$.  Now by the orthogonality of $|\Phi_0^{(1)}\>$ and
$|\Phi_1^{(1)}\>$ in \eqref{Schmidt}, 
\begin{equation}
\sum_{i_2, i_3, \ldots, i_{n}} {a_{i_2 i_3\cdots i_{n}}}^*
b_{i_2 i_3\cdots i_{n}} = 0\,,
\end{equation}
so, using the Cauchy-Schwarz inequality,
\begin{eqnarray}
|a_{11\cdots 1}|^2 |b_{11\cdots 1}|^2
& \leq & \left( \sum_{i_2,i_3,\ldots,i_{n}}
|a_{i_2 i_3\cdots i_{n}}|^2 - |a_{11\cdots 1}|^2\right)\nonumber\\
&&
\left( \sum_{i_2,i_3,\ldots,i_{n}}
|b_{i_2 i_3\cdots i_{n}}|^2 - |b_{11\cdots 1}|^2\right)
\nonumber \\
& = & (1-|a_{11\cdots 1}|^2)(1-|b_{11\cdots 1}|^2)\,.
\end{eqnarray}
Therefore
\begin{equation}
|a_{11\cdots 1}|^2 + |b_{11\cdots 1}|^2 \leq 1\,.
\end{equation}
Hence from Eq.\ (\ref{sum}) we have
\begin{equation}
\lambda_2 + \lambda_3 + \cdots + \lambda_n \geq A^2 = \lambda_1\,.
\end{equation}
Clearly there are similar inequalities with each of the $\lambda_k$ on
the right-hand side. We call these the \emph{polygon inequalities}.

To show that these inequalities define exactly the set of possible
one-qubit reduced states, we prove that given any real numbers $\{\lambda_1,\ldots
,\lambda_n\}$ lying between 0 and $1/2$ and satisfying \eqref{polygonk}
we can find an $n$-qubit state for which
$\lambda_k$ is the smaller eigenvalue of the reduced density matrix
$\rho_k$. 
We will suppose that $\lambda_1$ is the largest of the
numbers, so that
\begin{equation}\label{order}
0 \le \lambda_i \le \lambda_1 \le 1/2, \quad i=2,3,\ldots n.
\end{equation}

For $n = 3$ let
\begin{equation}
|\Psi_3\rangle \equiv a|100\rangle + b|010\rangle + c|001\rangle
+ d|111\rangle\,,
\end{equation}
where $a$, $b$, $c$ and $d$ are real and satisfy
$a^2 + b^2 + c^2 + d^2 = 1$. Then for each of the three qubits the
eigenvectors of the one-qubit reduced state are $|0\>$ and $|1\>$. 
Let $\lambda_i$ be the eigenvalue corresponding
to $|0\rangle$ for the $i$th qubit.  Then
\begin{eqnarray}
\lambda_1 & = & b^2 + c^2\,,\\
\lambda_2 & = & c^2 + a^2\,,\\
\lambda_3 & = & a^2 + b^2\,.
\end{eqnarray}
These can be solved as
\begin{eqnarray}
a^2 & = & \half\(\lambda_2 + \lambda_3 - \lambda_1\)\,, \label{a2} \\
b^2 & = & \half\(\lambda_3 + \lambda_1 - \lambda_2\)\,,\label{b2} \\
c^2 & = & \half\(\lambda_1 + \lambda_2 - \lambda_3\)\,. \label{c2}
\end{eqnarray}
Thus if the $\lambda_i$ satisfy the triangle inequalities there is a real
solution $(a,b,c)$ of the equations, and if $\lambda_i \le 1/2$ this
solution satisfies
\begin{equation}
a^2 + b^2 + c^2 = \frac{\lambda_1 + \lambda_2 + \lambda_3}{2} \leq 3/4\,,
\end{equation}
so that $a^2 + b^2 + c^2 \leq 1$ and it is possible to find a normalised
state $|\Psi\>$.  Hence, 
there is a state with arbitrary eigenvalues satisfying
Eqs.\ (\ref{polygon}) and (\ref{order}) for $n=3$.

Now assume that this is true 
for any $(n-1)$-qubit system.  Suppose that $\lambda_n$ is the smallest
of the $\lambda_i$, and 
define $\Lambda_1 \equiv \lambda_1 - \lambda_n$. There are two cases to consider.
If $\Lambda_1 \geq \lambda_i$ for $2 \leq i \leq n-1$, then
$(\Lambda_1,\lambda_2,\ldots,\lambda_{n-1})$ satisfies Eqs.\ (\ref{polygon})
and (\ref{order}) with $n$ replaced by $n-1$ and $\lambda_1$ by $\Lambda_1$.
In the second case, $\Lambda_1 < \lambda_m$ for some $m$, say $m=2$; we
can suppose that
$\lambda_2 \geq \lambda_i$ for $i=3,\ldots,n-1$.  Then
\begin{align*}
\Lambda_1 + \lambda_3 + \cdots + \lambda_{n-1} 
&= \lambda_1 +(\lambda_3-\lambda_n) + \lambda_4 + \cdots + \lambda_{n-1}\\
&\geq \lambda_1 \geq \lambda_2\,.
\end{align*}
Hence the set $(\lambda_2,\Lambda_1,\lambda_3,\ldots,\lambda_{n-1})$ satisfies
Eqs.\ (\ref{polygon}) and (\ref{order}) with $n$ replaced by $n-1$ and 
$\lambda_1$ and $\lambda_2$ replaced by $\lambda_2$ and $\Lambda_1$, 
respectively.  In either
case there is a state for which $\Lambda_1$, $\lambda_2$, $\ldots$,
$\lambda_{n-1}$ are the smaller eigenvalues of the reduced density matrices.  Let this state be
\begin{equation}
|\Psi_{n-1}\rangle = |0\rangle |\phi\rangle
+ |1\rangle  |\psi\rangle\,,
\end{equation}
where $|\phi\>$ and $|\psi\>$ are $(n-2)$-qubit states satisfying 
$\langle \phi\,|\,\phi\rangle = \Lambda_1$,
$\langle \psi\,|\,\psi\rangle = 1-\Lambda_1$ and 
$\langle \phi\,|\,\psi\rangle = 0$, and 
$\lambda_2$, $\lambda_3$,
$\ldots$, $\lambda_{n-1}$ are the smaller eigenvalues of the 
one-qubit reduced density matrices of
$|\phi\rangle \langle \phi| + |\psi\rangle \langle \psi|$.
Now consider the following $n$-qubit state:
\begin{equation}
|\Psi_n\rangle = |0\rangle |\phi \rangle 
|1\rangle + \sin\chi |0\rangle  |\psi \rangle 
|0\rangle + \cos\chi |1\rangle |\psi \rangle 
|1\rangle.
\end{equation}
The smaller eigenvalue $\tilde{\lambda}_i$ of this state
for the $i$th qubit with $2\leq i \leq n-1$ is again
the smaller eigenvalue of the $i$th one-qubit reduced density matrix of 
$|\phi\rangle \langle \phi| + |\psi\rangle\langle \psi|$.  Hence 
$\tilde{\lambda}_i = \lambda_i$ for $2 \leq i \leq n-1$.  
The $n$th and $1$st eigenvalues corresponding to the eigenvectors $|0\>$ are
\begin{eqnarray}
\tilde{\lambda}_n & =  & 
\sin^2\chi \langle \psi\,|\,\psi\rangle\,,\\ 
\tilde{\lambda}_1 & =  & \langle \phi\,|\,\phi\rangle + \sin^2\chi 
\langle \psi\,|\,\psi\rangle = \Lambda_1 + \tilde{\lambda}_n\,.\label{lambda1}
\end{eqnarray}
Since $\langle \psi\,|\,\psi\rangle \geq 1/2$, one can choose $\chi$ so
that $\tilde{\lambda}_n = \lambda_n\,.$  
Then \eqref{lambda1} gives 
$\tilde{\lambda}_1 = \lambda_1$. 

We conclude by induction that a state with the required one-qubit eigenvalues exists
for any $n$.

Of course, it does not follow that if an $n$-qubit state has one-qubit
reduced states whose eigenvalues satisfy the polygon inequalities, then
the $n$-qubit state must be pure. In general, a given set of one-qubit
reduced states can be obtained from many different $n$-qubit states,
most of which will be mixed: for example, the tensor
product of the one-qubit states, which will be mixed if any of the
one-qubit states are mixed. It is only when all the one-qubit states are
pure that the $n$-qubit state from which they arise must be pure.

Purity of the $n$-qubit state places no restriction on the eigenstates of
the one-qubit reduced states, since any pair of orthogonal eigenstates can
be transformed to any other by local unitary operations. The complete set of
one-qubit reduced states which can arise from a pure $n$-qubit state 
is therefore determined by the set of possible
eigenvalues that we have described. The situation for the reduced states
of larger subsets is likely to be more complicated. The example of
two-qubit subsets of a system of four qubits shows that
polygon inequalities \eqref{polygon} for any combination of eigenvalues 
are not sufficient, for these would allow all reduced states to be
totally mixed (all eigenvalues equal to 1/4), which is not possible
\cite{bobcarol}. It would be interesting to know the exact
set of marginal two-qubit probabilities for this system.

The situation for qudits (particles whose state spaces have dimension
$d$) also appears to be more complicated. An argument similar to the
above shows that the eigenvalues of the one-qudit reduced states must
satisfy \eqref{polygon} with each $\lambda_i$ replaced by the sum of all
but the largest eigenvalue of qudit $i$ (so the largest eigenvalues
satisfy the same inequality as for qubits), but there appear to be further inequalities
that must be satisfied. This is under investigation.

\end{document}